\documentstyle[12pt]{article}

\input mssymb			% only necessary for integers defn
\def\integers{\Bbb Z}

\advance\voffset by -2.2cm
\advance\hoffset by -2cm
\def\eq{\begin{equation}}
\def\eqe{\end{equation}}
\def\eqa{\begin{eqnarray}}
\def\eqae{\end{eqnarray}}

\def\half{\frac{1}{2}}

\def\cplex{{\mathchoice {\setbox0=\hbox{$\displaystyle\rm C$}\hbox{\hbox
to0pt{\kern0.4\wd0\vrule height0.9\ht0\hss}\box0}}
{\setbox0=\hbox{$\textstyle\rm C$}\hbox{\hbox
to0pt{\kern0.4\wd0\vrule height0.9\ht0\hss}\box0}}
{\setbox0=\hbox{$\scriptstyle\rm C$}\hbox{\hbox
to0pt{\kern0.4\wd0\vrule height0.9\ht0\hss}\box0}}
{\setbox0=\hbox{$\scriptscriptstyle\rm C$}\hbox{\hbox
to0pt{\kern0.4\wd0\vrule height0.9\ht0\hss}\box0}}}}

\def\reals{{\mathchoice {\setbox0=\hbox{$\displaystyle\rm R$}\hbox{\hbox
to0pt{\kern0.4\wd0\vrule height0.9\ht0\hss}\box0}}
{\setbox0=\hbox{$\textstyle\rm R$}\hbox{\hbox
to0pt{\kern0.4\wd0\vrule height0.9\ht0\hss}\box0}}
{\setbox0=\hbox{$\scriptstyle\rm R$}\hbox{\hbox
to0pt{\kern0.4\wd0\vrule height0.9\ht0\hss}\box0}}
{\setbox0=\hbox{$\scriptscriptstyle\rm R$}\hbox{\hbox
to0pt{\kern0.4\wd0\vrule height0.9\ht0\hss}\box0}}}}

\def\del{\partial}

\def\a{\alpha}

\def\d{\delta}
\def\e{\epsilon}		% Also, \varepsilon
\def\l{\lambda}
\def\m{\mu}
\def\n{\nu}
\def\w{\omega}
\def\s{\sigma}			%	\varsigma

\def\z{\zeta}

\def\S{\Sigma}

\def\cm{{\cal M}}
\def\cv{{\cal V}}
\def\ct{{\cal T}}

\def\zb{\bar{z}}

\begin{document}
\baselineskip 16pt plus 1pt minus 1pt
\hfill DAMTP 94-8
\vskip 24pt
\centerline{\bf Vortex scattering at near-critical coupling}
\vskip 24pt
\centerline{P.A.Shah}
\vskip 12pt
\centerline{\it Department of Applied Mathematics and Theoretical Physics}
\centerline{\it University of Cambridge}
\centerline{\it Silver Street, Cambridge CB3 9EW, U.K.}
\vskip 50pt
\centerline{\bf Abstract}
\vskip 24pt
The scattering of vortices at a critical value of the coupling
constant in the Lagrangian can be approximated by
a geodesic motion in the moduli space of classical static configurations of
vortices. In this paper we give a scheme for generalising this idea to
couplings that are near to the critical value. By perturbing a
critically coupled field, we show that scattering of vortices at
near-critical coupling can be approximated by motion in the original
moduli space with a perturbed metric, and a potential.
We apply this method to the scattering of two vortices, and compare
our results to recent numerical simulations, and find good agreement
where the scattering is not highly sensitive to radiation into other
field modes. We also investigate the possibility of bound stable
orbits of two vortices in the quantum field theory.

\vfill
\vskip -12pt
\hfill February 1994
\eject

\section{Introduction}
Solitons in two and three space dimensions are well-known products of
theories with spontaneously broken symmetries. Examples are vortices
in two dimensions, and cosmic strings, monopoles and skyrmions in three
dimensions \cite{solitons}. Individual solitons are solutions of the classical
field equations, with a localised energy density and conserved
topological charge.
\par
A difficult task is to determine the behaviour of soliton-soliton
interactions, especially for theories where the solutions of the field
equations cannot be written in closed form, but must be computed
numerically. One approach that has attracted study is to put the field
equations on a lattice, and to simulate the field theory on a
computer. This approach has been fruitful in studying the interactions
of vortices and strings \cite{stringsims} , and monopoles and
skyrmions to a lesser extent, but is however limited to
describing the behaviour for definite values of parameters of the
theory, and cannot provide a general description of soliton scattering.
\par
Another approach has been to examine the configuration
space of the field theory (or rather, the ``true'' configuration space
with gauge transformations quotiented out). The aim is to determine a
finite dimensional subset of this space on which the field equations
can be approximated by a simpler particle-like dynamical system.
Motion perpendicular to this space is ignored in the expectation
that it doesn't have a significant effect on the dynamics for
low velocities in the scattering. The scattering of monopoles
\cite{templeraston} and vortices \cite{samols} has been
modelled by adiabatic motion through
the space of static configurations of the field theory at fixed
topological charge, known as the {\it moduli space}.
In the case of monopoles and vortices,
current theory has only been able to model the interactions at a
certain value of a dimensionless coupling constant in the Lagrangian,
known as the {\it critical coupling}. This is known as the
moduli space approximation, and was first proposed by
Manton \cite{geoapprox}.
\par
This paper describes how the moduli space approximation can be
extended to describe soliton scattering at near-critical coupling.
Although our results are generalisable to other theories, we
concentrate on the scattering of vortices in the
Abelian Higgs model in $(2+1)$ dimensions. This is a generic
example of soliton scattering and describes the interactions of
straight, parallel $U(1)$ cosmic strings or magnetic flux tubes in a
superconductor in the Landau-Ginzburg model. Sec. 2 outlines the
theory of the moduli space approximation at critical coupling,
and its extension to near-critical coupling. Sec. 3 tests this
theory by applying it to the scattering of two vortices, and
compares the results with numerical simulations of the full field
theory. We conclude with a summary of our results.

\section{Extending the moduli space approximation}

We work with the Abelian Higgs model, with the vortices defined on
$\reals^{2}$, with a Lorentzian metric of signature $(+--)$.
\footnote{The vortices could also be defined on a general orientable
Riemann surface such as $S^{2}$ , $S^{1} \times S^{1}$ or
hyperbolic space $H^{2}$. We take $\reals^{2}$ here for simplicity.}
The model has spontaneously broken $U(1)$ gauge symmetry, and action
\eq
\label{eq:action}
S = \int d^{3}x [ -\frac{1}{4}f_{\m\n}f^{\m\n} + \half D^{\m}\phi
(D_{\m}\phi)^{*} + \a (|\phi|^{2} - c_{0}^{2})^{2}]
\eqe
where $\phi$ is complex scalar valued. $D_{\m}\phi =
(\del_{\m} -ieA_{\m})\phi$ where $A_{\m}$ is the real valued gauge
potential, and $e$ the gauge coupling. The action (\ref{eq:action})
can be rescaled using the dimensionless quantities
\eqa
x^{\m} & = & \frac{1}{ec_{0}}\tilde{x}^{\m} \\
\phi & = & c_{0}\tilde{\phi} \nonumber \\
A_{\m} & = & c_{0}\tilde{A_{\m}} \nonumber \\
S & = & \frac{c_{0}}{e} \tilde{S} \nonumber
\eqae
to read
\eq
\tilde{S} = \int d^{3}\tilde{x} [ -\frac{1}{4}\tilde{f}_{\m\n}
\tilde{f}^{\m\n} +
\half \tilde{D}^{\m}\tilde{\phi} (\tilde{D}_{\m}\tilde{\phi})^{*} +
(\frac{1}{8} +\l)(|\tilde{\phi}|^{2} - 1)^{2}]
\eqe
where $\frac{1}{8} +\l = \frac{\a}{e^{2}}$ (so $\l > -\frac{1}{8}$)
and $\tilde{D_{\m}}\tilde{\phi} = (\tilde{\del_{\m}} -
i\tilde{A_{\m}})\tilde{\phi}$. In these units the mass of the gauge
particle $m_{photon} = 1$, and the mass of the scalar field is
$m_{Higgs} = \sqrt{1+8\l}$. We work with this action, dropping the
tildas. The action can be split into kinetic and potential pieces,
which after choosing the gauge $A_{0}=0$ can be written
\eqa
S & = & \int dx_{0} ( \ct - \cv) \label{eq:ct} \\
\ct & = & \int d^{2}x \half ( \dot{\phi} \dot{\phi}^{*} +
\dot{A_{i}} \dot{A_{i}}) \nonumber \\
\cv & = & \int d^{2}x \, [\half f_{12}^{2} + \half D_{i} \phi
(D_{i} \phi)^{*} + (\frac{1}{8} + \l)(|\phi|^{2} - 1)^{2} ] \nonumber
\eqae
(the equation of motion corresponding to $A_{0}$ must then be imposed
as a constraint; Gauss' Law). $\ct$ also provides a natural metric on
configuration space. Any finite energy configuration has
a magnetic flux
\eq
\int d^{2}x f_{12} = 2\pi N
\eqe
where $N$ is an integer, and this magnetic flux is localised in the
neighbourhoods of the $N$ zeros of $\phi$. This topological identity
can be used to write the potential energy as a sum of squares, plus a
remainder term. Introducing the complex coordinates $z=x_{1}+ix_{2}$,
$\bar{z} = x_{1}-ix_{2}$,
\eq
\label{eq:cv}
\cv  =  \pi N  + \int \frac{dz \wedge d\zb}{2i}
[\half (f_{12} + \half (|\phi|^{2} - 1))^{2} +
\half | D_{\zb} \phi | ^{2} + \l (|\phi|^{2} - 1)^{2}]
\eqe
At $\l=0$ the last, remainder, term in the above equation vanishes.
The potential energy $\cv$ can then saturate its topological
lower bound $N\pi$ when the Bogomolny equations \cite{bogoeqns}
\eqa
\label{eq:bogoeqns}
D_{\bar{z}}\phi & = & 0 \\
f_{12} + \half(|\phi|^{2} - 1) & = & 0 \nonumber
\eqae
are satisfied ($\l=0$ is the critical coupling for this theory). The
first equation of (\ref{eq:bogoeqns}) can be solved to give
\eq
A_{\zb} = -i \del_{\zb} \log{\phi}
\eqe
so that the gauge field is dependent on the Higgs field, and can be
eliminated as a variable in the equations (\ref{eq:bogoeqns}).
When $\ct=0$ and $\cv=N\pi$, these equations give the
static solutions of the field equations
$(\phi^{(0)}(Q,z,\zb), A^{(0)}_{i}(Q,z,\zb))$. $Q$ is a set of
collective coordinates representing the position of the field in the
moduli space. This is because at critical coupling there are no
forces between static vortices. It has been shown \cite{bradlow} that
on any orientable Riemann surface, $Q$ can be specified by the
unlabelled set of the zeros of $\phi^{(0)}$, $\{ z_{1} \ldots z_{N} \}$,
which may be identified with the positions of the vortices.
An important aspect of soliton behaviour is that solitons are
indistinguishable particles at the classical
level, and so the coordinates $Q$ are invariant under the action of
the permutation group $\S_{N}$ which exchanges vortex positions. The
moduli space of $N$ vortices defined on an orientable
Riemann surface $M$ is then $\cm_{N} = \frac{M^{N}}{\S_{N}}$.
\par
The essence of the moduli space approximation at critical coupling
is to project the dynamic field evolution onto the moduli space.
The projection $\phi \rightarrow \phi^{(0)}(Q)$ is made by
fixing the zeros of the Higgs field, and the dynamics of the
field are approximated by allowing the coordinates $Q$ to
vary in time. The time derivative of the Higgs field
is approximated by
\eq
\dot{\phi}^{(0)} = \lim_{\d t \rightarrow 0}
\frac{\phi^{(0)}(Q(t+\d t)) - \phi^{(0)}(Q(t))}{\d t}
\label{eq:kinapprox}
\eqe
and similarily for the gauge field $A_{i}$.
Samols \cite{samols} has shown that, using the approximation
(\ref{eq:kinapprox}), the kinetic energy of the fields can be
written as
\eqa
\label{eq:samolst}
\ct & = & \sum_{i,j=1}^{N} \half\pi g_{ij}(Q) \frac{dz_{i}}{dt}
\frac{d\zb_{j}}{dt} \\
g_{ij} & = & \d_{ij} + 2\frac{\del \bar{b}_{j}}{\del z_{i}}
\nonumber
\eqae
where $\bar{b}_{j}$ is the linear coefficient in an expansion of
$f(z,\zb)=\log{|\phi^{(0)}|^{2}}$ about the vortex at $z_{j}$:
\eqa
f(z,\zb) & = & \log{|z-z_{j}|^{2}} + a_{j}(Q) + \half b_{j}(Q)(z-z_{j}) +
\half \bar{b}_{j}(Q)(\zb-\zb_{j}) + \\
& & c_{i}(Q)(z-z_{j})^{2} + \bar{c}_{j}(Q)(\zb-\zb_{j})^{2}
+ d_{j}(Q)(z-z_{j})(\zb-\zb_{j})
+ \ldots
\nonumber \eqae
When there are no massless fields
present, as in this theory, motion perpendicular to the moduli space
is suppressed by the positive quadratic nature of the potential
in the directions orthogonal to $\cm_{N}$. The normal frequencies are
bounded below by the photon mass and Higgs mass.
Whilst no rigorous field theory
analysis has been done, finite dimensional theory suggests that in a
scattering process where the timescale of the process along the moduli
space is $t_{1}$ and the timescale perpendicular to the moduli space
is $t_{2}$, the fraction of energy transferred to modes perpendicular
to the moduli space is $O(\exp{-t_{1}/t_{2}})$. For vortex
scattering with typical velocity $v$, this is $O(\exp{-\frac{1}{v}})$
and so is suppressed at small $v$. This approximation has been
tested by computer simulations, and found to work well up to
velocities $v \simeq 0.4$ \cite{samols}.
\par
Expressing the motion of solitons in terms of a drift of a
projection of the field onto the moduli space of static configurations
suggests a similar approach when $\l \neq 0, \l$ small. When
$\l \neq 0$, the vortices exert a static force on each other, and
the space of static solutions collapses to isolated points in
configuration space when spatial translations and rotations are
quotiented out (for example, all $N$ vortices coalesce into a stable
winding number $N$ vortex in the attractive case). One possible
way to model the interactions of weakly attracting/repelling vortices
would be to project the motion onto the set of curves of steepest descent
of the potential, connecting unstable to stable static configurations. This
technique originates in molecular reaction dynamics, and might be
applied to the scattering of skyrmions \cite{skyrmionscat}. The true
evolution is expected to follow a path in some neighbourhood of these
gradient curves, for sufficiently small kinetic energies. However, the
calculation of these gradient curves is a difficult constrained
optimisation problem. In this paper, we argue that the
dynamics is equally well approximated by projection onto a simpler,
nearby manifold: the moduli space at critical coupling.
\par
We conjecture that as the potential energy of a particular
configuration $\cv[\phi,A_{i},\l]$ changes smoothly with $\l$
according to equation (\ref{eq:cv}), the set of steepest
descent curves lies in the neighbourhood of the moduli space
of static configurations $(\phi^{(0)},A^{(0)}_{i})$ at critical
coupling (for sufficiently small $\l$), and
indeed is diffeomorphic to it. Although these latter configurations
are now no longer static solutions of the field equations
or of minimal potential energy, for $\l$ small enough
the evolution of the fields will remain in their neighbourhood.
\par
We propose an approximate description of the
scattering by projecting the field at any given time
to $\l=0$ as well as $\ct =0$,
keeping the zeros of the Higgs field fixed. The scattering is then
modelled by collective coordinate motion in the original
moduli space at critical coupling.
The induced metric on this space is a perturbed version of the metric
at critical coupling, however we give arguments below that it is not
necessary to accurately determine this perturbation for all
vortex positions. There is also an induced potential on the
moduli space. We may write
\eqa
\phi & = & \phi^{(0)}(Q(t)) \exp{\d h(z,\zb,t)} \label{eq:pert} \\
A_{i} & = & A^{(0)}_{i}(Q(t)) + \d B_{i}(z,\zb,t) \nonumber
\eqae
where $h$ and $B_{i}$ are $O(1)$, but $\d$ is small.
Taking $\e$ to be a measure of the (small) kinetic energy of the
fields, we assume that for $\l$ and $\e$ sufficiently small,
$\d (\l,\e)$ is $O(\l)$ small. There is actually a
technical complication with (\ref{eq:pert}) as $h$ and $B_{i}$ are not
independent, but are constrained by Gauss' Law. Furthermore, because
of the residual time-independent gauge symmetry, the solution may
evolve to be close to a {\it gauge transformed} multivortex.
We will not need to take account of this as a knowledge of $h$ and $B_{i}$
are not necessary for our approximation. The point of equation
(\ref{eq:pert}) is to estimate the nearness of the fields to the
moduli space at a particular time, and their kinetic energy.
Substituting (\ref{eq:pert}) into (\ref{eq:cv}) we find that the
potential
\eqa
\label{eq:potdef}
\cv & = & N\pi + \l \int d^{2}x (|\phi^{(0)}(Q)|^{2} - 1)^{2} +
O(\l^{2}, \l\d,\d^{2}) \\
    & \equiv & N\pi + \l V(Q) + O(\l^{2}) \nonumber
\eqae
is independent to first order in $\l$ of the deviation of the field from
$\cm_{N}$. This is
expected, because $\cm_{N}$ was at a minimum of the potential energy
for $\l=0$. Choosing a static configuration $\phi^{(0)}(Q)$ at $\l=0$
and ``turning on'' $\l$ results in the minimum of $\cv$ for fixed
$Q$ being at an $O(\l)$ distance from $\phi^{(0)}(Q)$,
but the difference between the new value of the potential of
$\phi^{(0)}(Q)$ given by (\ref{eq:potdef})
and the new minimum at fixed $Q$ is $O(\l^{2})$.
\par
The situation with the kinetic energy is not so simple; it
varies to first order in $\d$, as can be seen by direct substitution
of (\ref{eq:pert}) into (\ref{eq:ct}). To determine the first order
variation of the kinetic energy, we would in principle have to solve
for $h$ and $B_{i}$ by minimizing the potential energy at fixed $Q$.
However, we now give arguments that in a low velocity
scattering this is not necessary. It is not our intention to give a
rigorous proof of our intended approximation, but rather to base our
argument on the physical properties of vortex scattering.
\par
If we consider the scattering of near-critical vortices with typical
low velocity $v$
at a particular time, then the rate of angular deviation from the
critical coupling motion that can be attributed to the potential is
$O(\l /v)$. The potential is only appreciably different from zero
over the timescale of the scattering $t^{*} = O(1/v)$ so the
total angular deviation due to the potential can be estimated as
$O(\l /v^{2})$. The rate of
angular deviation from critical coupling motion that can be attributed
to the perturbation in the metric is $O(\l v^{2} / v)$. This acts for
all times, including before and after the scattering. But if we
can calculate the effect of the perturbed metric for times
$|t|>t^{*}$ when the vortices are well separated, and make
some approximation to it for the
times when the vortices are close then the effect of our lack of
knowledge of the metric perturbation when the vortices are close
is a total angular error of $O(\l)$. Thus for slow motion the effect of the
potential dominates over the perturbation in the metric, {\it
provided the perturbation is correct asymptotically}.
\par
To determine how the metric changes with $\l$ when the vortices are
well separated, let us first consider a static $N=1$ solution
$(\phi^{(\l)},A_{j}^{(\l)})$ at $\l \neq 0$.
This has potential energy which from (\ref{eq:potdef}) can be written as
\eqa
\cv & = & \pi + \l \int d^{2}x (|\phi^{(0)}|^{2} - 1)^{2} +O(\l^{2}) \\
  & \equiv & \pi + \l V_{1} + O(\l^{2}) \nonumber
\eqae
where $\phi^{(0)}$ is the Higgs field of a single $\l=0$ vortex.
$V_{1}$ is a constant, whose numerical value is given in Sec. 3. The
theory is Lorentz invariant, so we may obtain the energy of a moving
vortex by a Lorentz boost, which in the low velocity limit becomes a
Galilean boost. The energy is then
\eq
E= (\pi + \l V_{1} +O(\l^{2}))(1+\half v^{2} + O(v^{4}))
\eqe
so we may identify the potential energy of a single vortex as its
inertial mass
\eq
m = \pi +\l V_{1} +O(\l^{2})
\eqe
Consider now a configuration of slowly moving, well-separated
vortices at near-critical coupling, such that the significant
contribution to the kinetic energy comes from the motion of the
vortices (rather than radiative modes). The Higgs and gauge
fields are massive, and exponentially approach the vacuum
away from the vortices. Thus the significant part of the kinetic
energy comes from a region close to the zero of the Higgs field.
We assert that in this region, the configuration and its time
derivative is well approximated by the boosted $N=1$ solution
$(\phi^{(\l)},A_{j}^{(\l)})$ with the appropriate velocity, centred
at the zero of the Higgs field. The kinetic energy will then approach
the kinetic energy of $N$ moving particles of inertial
mass $m$
\eq
\label{eq:tapprox}
\ct \simeq \half m \sum_{i=1}^{N} \frac{dz^{i}}{dt} \frac{d\zb^{j}}{dt}
\eqe
This models the asymptotics of vortex dynamics. We therefore
choose as our approximation to the kinetic energy
\eq
\label{eq:kindef}
T(Q,\dot{Q}) = \sum_{i,j=1}^{N} \half (\pi +\l V_{1}) g_{ij}(Q)
\frac{dz^{i}}{dt} \frac{d\zb^{j}}{dt}
\eqe
where the metric $g_{ij}$ is defined in (\ref{eq:samolst}). As a
consequence of the exponential decay of static solutions,
the metric quickly becomes flat as the vortex separation increases,
and $T(Q,\dot{Q})$ quickly approaches (\ref{eq:tapprox}).
\par
In summary, the evolution of a configuration of vortices in a
scattering problem is expected to be well approximated by a
an evolution of particles (the zeros of the Higgs field) according to
the action
\eq
\label{eq:ourlag}
S = \int dt \; (T(Q,\dot{Q}) - \l V(Q))
\eqe
where $T(Q,\dot{Q})$ is defined in (\ref{eq:kindef}) and $V(Q)$ is
defined in (\ref{eq:potdef}).
\par
In a recent paper \cite{stuart}, Stuart has established some rigorous
results on vortex dynamics. An existence theorem is proved for an initial
value problem in configuration space, where the initial data is
$O(|\l|)$ close to the moduli space, the initial field
derivatives are $O(|\l|^{\half})$, and $O(|\l|)$ close to the
translational zero modes (at critical coupling) on the moduli space.
For a subsequent time $T^{*} = |\l|^{-\half}$ the evolution continues to
satisfy the same conditions as the initial data. The point on the
moduli space of closest approach to the field configuration is
$O(|\l|^{\half})$ close to a Lagrangian motion given by the
metric on the moduli space {\it at critical coupling}
(\ref{eq:samolst}) and the potential $V(Q)$ from (\ref{eq:potdef}).
This applies for all vortex configurations, not just for well
separated vortices. However, the metric used is
incorrect asymptotically, as the inertial mass of the vortices
changes with $\l$. This contributes to the
$O(|\l|^{\half})$ drift from the Hamiltonian evolution on the moduli
space (this can be seen immediately by considering the motion of a
single boosted near-critical vortex in the context of this problem).
We anticipate that Stuart's results would still hold if our metric
(\ref{eq:kindef}) was used together with the potential
to give the Lagrangian evolution
on the moduli space. Furthermore, in a scattering problem where the
vortices are well separated except for a time of $O(T^{*})$, we
conjecture that Stuart's results may be improved to hold for all times
with the use of our asymptotic metric.
\par
We now seek to test the validity of our approximation
against ``observation'': numerical lattice simulations of the
full field theory.

\section{The metric and potential of two vortices}

For two vortices, the geometry of the moduli space is very simple.
Using the notation of Samols, we decompose the moduli space as the product
\eq
M_{2} = \cplex \times M_{2}^{0}
\eqe
where, if vortices with masses $m$ lie at $z_{1}$ and $z_{2}$ then
the centre of mass coordinate $Z=\half(z_{1}+z_{2}) \in \cplex$
can be quotiented out, leaving the system with a reduced mass of
$m/2$ and relative
coordinate $\z = \half(z_{1}-z_{2}) \in M_{2}^{0}$. The vortices lie
at $\pm \z$ in these coordinates. We will find it convenient to use
the polar coordinates
\eq
\z= \s e^{i\theta}
\eqe
where, due to the symmetry of the scattering under interchange of two
vortices, the range of $\theta$ is $[0,\pi)$. Hermiticity of the
metric on $M_{2}^{0}$ (in addition to rotational symmetry)
follows from the reality of the kinetic energy,
and requires it to have the form
\eq
ds^{2} = F^{2}(\s)(d\s^{2} + \s^{2}d\theta^{2})
\eqe
$F^{2}(\s)$ can be computed from the solution $\phi^{(0)}(\s)$
of the Bogomolny equations with zeros of $\phi^{(0)}$ on the real axis
at $\z=\pm \s$ by the formula
\eq
F^{2}(\s) = 1+ \frac{1}{\s}\frac{d}{d\s}(\s b)
\eqe
where $b$ is the coefficient in the expansion of
$\log{|\phi^{(0)}(\s)|^{2}}$ around the vortex at $\z = \s$ (see
(\ref{eq:samolst})).
As explained by Samols, the two dimensional space $M_{2}^{0}$
has the form of a rounded cone when embedded in Euclidean $\reals^{3}$.
\par
The potential on the moduli space away from critical coupling
is clearly a function of $\s$ only. Recalling (\ref{eq:potdef})
it is calculated by
\eq
\label{eq:twovortpot}
V(\s) = \int_{\reals^{2}} (|\phi^{(0)}(\s)|^{2}-1)^{2} d^{2}x
\eqe
The interactions between vortices are smooth when expressed in terms
of a good global coordinate. Due to the symmetry of the field under
the interchange of two vortices, $\z$ is not well-defined as $\z
\rightarrow 0$, but $\w=\z^{2} \in \cplex$ is. For small $\s$
the leading order behaviour of $V(\s)$ is determined by the
requirement that $\frac{dV(\w)}{d\w}$ exists at $\w=0$; in fact by
rotational symmetry, $\frac{dV}{d\w} = 0$ at $\w=0$. If we define
$V_{2} = V(0)$, $V(\w)=V_{2}+O(\w^{2})$ and
\eq
V(\s) = V_{2}+\a\s^{4}+O(\s^{6}) \; \mbox{for} \; \s \ll 1
\eqe
By numerical computation, $V_{2} \simeq 13.63$ and $\a \simeq 0.36$.
For large vortex separation, the contribution to the integral
(\ref{eq:twovortpot}) from the neighbourhood of one vortex is only
slightly perturbed by the presence of the other. For the Higgs
field $\phi^{(0)}_{\s}$ of a single vortex at $\z=\s$
\eq
0 < 1-|\phi^{(0)}_{\s}|^{2}< M(\d) e^{-(1-\d)\s} \; \; \mbox{for any}
\; \; \d > 0 , |z-\s|>\s
\eqe
which implies
\eq
|\phi^{(0)}(\s)|^{2} = |\phi^{(0)}_{\s}|^{2}|\phi^{(0)}_{-\s}|^{2} +
O(e^{-2(1-\d)\s})
\eqe
so
\eq
2V_{1} < V(\s) < 2V_{1} + O(e^{-(1-\d)\s})
\eqe
Numerically, $2V_{1} \simeq 10.44 < V_{2}$.
\footnote{This argument can be generalised to $N$ vortices.
If $\{z_{1} \ldots z_{N} \}$ are the vortex positions, then $NV_{1} < V
< NV_{1} + O(e^{-(1-\d)\s})$ where here
$\s = min_{i \ne j}(|z_{i}-z_{j})|$. $NV_{1} < V_{N} = V(0)$}
\par
The numerical computation of the metric and potential had two stages.
Firstly, the field $\phi^{(0)}(\s)$ was calculated from the Bogomolny
equations using an initial approximation which was a
superposition of two approximate 1-vortex solutions. This was relaxed using a
multigrid algorithm on a Silicon Graphics Indigo workstation. Seven
grids were used, the densest of which was $256 \times 384$.
Convergence was typically quite rapid, requiring only a few sweeps
over the range of grids. The derivative of $\ln |\phi^{(0)}(\s)|^{2}$ (for
the metric) at the vortex
position was calculated by fitting a cubic spline curve along the real
axis joining the two vortices. The integral $V(\s)$ given by equation
(\ref{eq:twovortpot}) was performed by fitting
bicubic surface splines to the lattice solution. This was done for
values of $\s$ from $0$ to $8$ at intervals of $0.02$. Finally,
a suitably smoothed spline curve was fitted to the functions
$F^{2}(\s)$ and $V(\s)$. A graph of the potential $V$ is shown in Fig.
\ref{fig:potential}. Jacobs and Rebbi \cite{jacobs} have calculated the
interaction energy of two stationary vortices at coupling constants (in
our units) of $\l=0.08625$ and $\l=-0.06375$. They use a functional ansatz
for the Higgs field which is a weighted superposition of the Higgs
field for two isolated vortices at $\pm \s$ and a 2-vortex at the
origin, plus a correction term which is expanded as a power series in
$|z|$ and cos(arg $z$). The free parameters are the coefficients of
the power series which is truncated at sixth order. They then
optimise the potential energy numerically as a function of
those parameters. A scaled comparison of their results with
our potential is shown in Fig. \ref{fig:jr.pot.compare.norm}.

%ffffffffffffffffffffffffffffffffffffffffffffffffffffffffffffffffffffffff
\begin{figure}
\hspace{2cm}
%%%\epsfxsize=13cm
%%%\epsfbox{potential.eps}
\caption{A plot of $V(\s)$ for two vortices where the
inter-vortex separation is $2\s$.}
\label{fig:potential}
\end{figure}
%fffffffffffffffffffffffffffffffffffffffffffffffffffffffffffffffffffffffff

%ffffffffffffffffffffffffffffffffffffffffffffffffffffffffffffffffffffffff
\begin{figure}
\hspace{2cm}
%%%\epsfxsize=13cm
%%%\epsfbox{jr.pot.compare.norm.eps}
\caption{A comparison of the stationary vortex interaction
potential calculated by Jacobs and Rebbi for $\l=0.08625$ (upper
curve) and $\l=-0.06375$ (lower curve), scaled to our units,
to the potential $V(\s)-V(\infty)$ (dashed line)}
\label{fig:jr.pot.compare.norm}
\end{figure}
%fffffffffffffffffffffffffffffffffffffffffffffffffffffffffffffffffffffffff

\section{Scattering behaviour of two vortices}
The scattering of two vortices at critical coupling has for some time
been studied numerically by putting the field theory on a
lattice \cite{stringsims}. Recently, Myers, Rebbi and Strilka \cite{MRS}
simulated vortex scattering on a lattice at couplings $\l
=-\frac{1}{16}, 0 , \frac{1}{8}$ (in their units, this corresponded to
$\l =1,2,4$) \cite{MRS}. Here, we compare our results at the values of
the coupling of $\l=-\frac{1}{16}$ and $\l = \frac{1}{8}$ with theirs.
For a comparison of $\l=0$ scattering with numerical simulation, see
ref. \cite{samols}.
\par
We turn first to $\l=\frac{1}{8}$ (type-II) vortices.
The vortices were initially placed at large x-separations on paths with
velocities $(\pm v,0)$ and an impact parameter of $2d$. The equations
of motion derived from our model Lagrangian (\ref{eq:ourlag})
were integrated numerically using a fourth order Runge-Kutta
method with variable step size. In Fig.
\ref{fig:vortex.path.plus} we show some trajectories for
initial vortex velocity of $v=0.3$.

%ffffffffffffffffffffffffffffffffffffffffffffffffffffffffffffffffffffffff
\begin{figure}
\hspace{1.5cm}
%%%\epsfxsize=14cm
%%%\epsfbox{vortex.path.plus.eps}
\caption{Some sample trajectories in the centre of mass frame
for vortices with $\l=\frac{1}{8}$
and initial velocity $v=0.3$. The lowest impact parameter graph is at
$d=0.01$. The trajectory of only one vortex is shown.}
\label{fig:vortex.path.plus}
\end{figure}
%fffffffffffffffffffffffffffffffffffffffffffffffffffffffffffffffffffffffff

Scattering angles for a range of impact parameters and velocities are
shown in Table 1. As the interaction is repulsive, there is a
critical velocity below which a head on collision will scatter the
vortices elastically back through $180^{\circ}$, and above which the
vortices will scatter through $90^{\circ}$. This is calculable from
the conservation of energy in our approximation, giving
\eq
v_{crit} = \sqrt{\frac{\l(V(0)-V(\infty))}{2\pi}} \simeq 0.71 \sqrt{\l}
\eqe
so for $\l=\frac{1}{8}$, $v_{crit} \simeq 0.25$. For initial
velocities below $v_{crit}$ the scattering angle increases smoothly
with decreasing impact parameter. For velocities above, but close to
$v_{crit}$ the scattering angle has a turning point at very small impact
parameter, and scattering in the backwards direction is still possible. As
the velocity increases further, this turning point disappears, and all
scattering is in the forward direction. A comparison with the results
of Myers et al.  is shown in Fig. \ref{fig:compare.deflect.plus}.
The agreement
is very good at low velocities, but less good as the velocity
increases. This is due to the non-relativistic nature of the
moduli space approximation, and radiation into other field modes.
This loss of quantitative accuracy is observed to a similar
degree in the moduli space approximation for critical vortices.
\par

%ttttttttttttttttttttttttttttttttttttttttttttttttttttttttttttttttttttttt
\begin{table}
\vspace{0.5cm}
\caption{The scattering angle $\theta_{defl}$ for
$\lambda=\frac{1}{8}$ is plotted versus impact parameter and initial
velocity $v$.}
\begin{center}
\begin{tabular}{|l|rrrrrrrrrr|}
\hline \hline
 & \multicolumn{10}{c|}{impact parameter, $d$} \\
$v =$ &
0.5   &   1.0 &   1.5 &   2.0 &   2.5 &   3.0 &   3.5 &   4.0 &  4.5 &  5.0 \\
\cline{2-11}
0.05 &
160.1 & 140.3 & 120.6 & 101.2 &  82.3 &  64.0 &  46.9 &  31.7 & 19.3 & 10.5 \\
0.1 &
151.7 & 124.7 &  99.5 &  76.6 &  56.1 &  38.5 &  24.2 &  13.7 &  7.0 & 3.3 \\
0.15 &
140.3 & 106.9 &  80.0 &  57.7 &  39.3 &  24.7 &  14.1 &   7.2 &  3.5 & 1.5 \\
0.20 &
119.1 &  87.2 &  63.5 &  44.2 &  28.7 &  17.1 &   9.2 &   4.5 &  2.1 & 0.9 \\
0.25 &
 90.4 &  70.8 &  51.8 &  35.4 &  22.2 &  12.6 &   6.6 &   3.1 &  1.4 & 0.6 \\
0.30 &
 77.3 &  61.0 &  44.3 &  29.6 &  18.0 &   9.9 &   5.0 &   2.3 &  1.0 & 0.4 \\
0.35 &
 72.6 &  55.7 &  39.5 &  25.7 &  15.2 &   8.1 &   4.0 &   1.8 &  0.8 & 0.3 \\
0.40 &
 70.5 &  52.6 &  36.4 &  23.1 &  13.2 &   6.9 &   3.3 &   1.5 &  0.6 & 0.3 \\
0.45 &
 69.4 &  50.7 &  34.3 &  21.2 &  11.8 &   6.0 &   2.9 &   1.3 &  0.5 & 0.2 \\
0.50 &
 68.7 &  49.5 &  32.9 &  19.8 &  10.8 &   5.4 &   2.5 &   1.1 &  0.5 & 0.2 \\
\hline \hline
\end{tabular}
\end{center}
\end{table}
%tttttttttttttttttttttttttttttttttttttttttttttttttttttttttttttttttttttttt

%ffffffffffffffffffffffffffffffffffffffffffffffffffffffffffffffffffffffff
\begin{figure}
\hspace{2.0cm}
%%%\epsfxsize=13cm
%%%\epsfbox{compare.deflect.plus.eps}
\caption{A comparison of scattering angles calculated by Myers et al.
to values calculated using the moduli space approximation when
$\l=\frac{1}{8}$.
The impact velocities shown are $v=0.1$ (+), $v=0.2$ (o),
$v=0.3$ (x), $v=0.4$ ($\ast$)}
\label{fig:compare.deflect.plus}
\end{figure}
%fffffffffffffffffffffffffffffffffffffffffffffffffffffffffffffffffffffffff

Turning now to $\l=-\frac{1}{16}$ vortices (type-I), the
attraction due to the potential is longer range than the repulsion due
to the metric (the mass of the photon is larger than the mass of the
Higgs particle). Scattering trajectories for initial velocity $v=0.3$
are shown in Fig. \ref{fig:vortex.path.minus}, and Table 2 gives
scattering angles for a
range of initial velocities and impact parameters.
The scattering behaviour is now qualitatively the same at all
velocities. For large impact
parameter, the scattering angle is small and negative. As the impact
parameter decreases, the scattering angle decreases to a minimum, and
then as the repulsion of the vortex cores overcomes the attraction of
the potential, increases back through zero and approaches
$90^{\circ}$ scattering at zero impact parameter. Scattering at low
impact parameter is insensitive to the vortex velocity.  For initial
velocity $v=0.3$, and impact parameter $d \simeq 2.8$ the
repulsion of the vortex core balances the attraction of the
potential, giving zero deflection.

%ffffffffffffffffffffffffffffffffffffffffffffffffffffffffffffffffffffffff
\begin{figure}
\hspace{2.0cm}
%%%\epsfxsize=13cm
%%%\epsfbox{vortex.path.minus.eps}
\caption{Some sample trajectories in the centre of mass frame
for vortices with $\l=-\frac{1}{16}$
and initial velocity $v=0.3$. The lowest impact parameter graph is at
$d=0.01$. The trajectory of only one vortex is shown.}
\label{fig:vortex.path.minus}
\end{figure}
%fffffffffffffffffffffffffffffffffffffffffffffffffffffffffffffffffffffffff

%ttttttttttttttttttttttttttttttttttttttttttttttttttttttttttttttttttttttttt
\begin{table}
\vspace{0.5cm}
\caption{The scattering angle $\theta_{defl}$ for
$\lambda=-\frac{1}{16}$ is plotted versus impact parameter and initial
velocity $v$.}
\begin{center}
\begin{tabular}{|l|rrrrrrrrrr|}
\hline \hline
 & \multicolumn{10}{c|}{impact parameter, $d$} \\
$v =$ &
0.5  &  1.0 &  1.5 &  2.0 &   2.5 &   3.0 &   3.5 &   4.0 &  4.5 & 5.0 \\
\cline{2-11}
0.05 &
66.6 & 43.4 & 19.1 & -7.0 & -36.1 & -70.1 &-113.9 &-182.6 &-72.6 &-12.8 \\
0.1 &
63.7 & 37.1 & 8.8 & -22.9 & -60.7 &-109.4 & -61.0 & -16.9 & -6.2 & -2.5 \\
0.15 &
63.2 & 36.4 & 8.5 & -20.2 & -42.2 & -32.7 & -14.2 &  -5.8 & -2.4 & -1.0 \\
0.20 &
63.6 & 37.5 & 11.9 & -10.4 & -19.5 & -13.4 & -6.5 &  -2.9 & -1.3 & -0.5 \\
0.25 &
64.0 & 39.0 & 15.4 &  -2.7 &  -9.2 &  -6.8 & -3.6 &  -1.7 & -0.7 & -0.3 \\
0.30 &
64.5 & 40.2 & 18.2 &   2.1 &  -4.1 &  -3.7 & -2.1 &  -1.0 & -0.5 & -0.2 \\
0.35 &
64.9 & 41.2 & 20.1 &   5.2 &  -1.2 &  -1.9 & -1.2 &  -0.6 & -0.3 & -0.1 \\
0.40 &
65.2 & 42.0 & 21.5 &   7.2 &   0.7 &  -0.8 & -0.7 &  -0.4 & -0.2 & -0.1 \\
0.45 &
65.4 & 42.6 & 22.6 &   8.6 &   1.9 &  -0.1 & -0.3 &  -0.2 & -0.1 & -0.1 \\
0.50 &
65.6 & 43.0 & 23.4 &   9.6 &   2.7 &   0.4 & -0.1 &  -0.1 & -0.1 & -0.0 \\
\hline \hline
\end{tabular}
\end{center}
\end{table}
%tttttttttttttttttttttttttttttttttttttttttttttttttttttttttttttttttttttttt

\par
Interesting behaviour is observed at low velocity. The
largest (negative) scattering angle diverges as $v \rightarrow 0$;
the vortices
can orbit around each other, forming a ``glory'' (it has been shown
\cite{gibbons} that this type of behaviour does not occur in the
scattering of photons off cosmic strings, where the photons
propagate in the background gravitational metric of the string).
Unfortunately, as the scattering angle
is very sensitive to changes in the impact parameter or velocity in
this region, it is also very sensitive to radiation into other
field modes, and so our approximation will only give a qualitative
description of the scattering. Myers et al. observed that for low impact
parameter and velocity, this radiation means the vortices can no
longer escape their attraction, and scatter repeatedly through $90^{\circ}$
with the maximum separation decreasing with each cycle; presumably
the vortices would eventually settle down to a stable winding number
two vortex.
We cannot observe this end result in our approximation, as the
conservation of energy and angular momentum forbids it,
but we do observe bound orbits in the two vortex system.
We turn to the question of their stability in the
next section. Figs. \ref{fig:compare.deflect.minus.1} and
\ref{fig:compare.deflect.minus.2} give a comparison our results
to Myers et al.
Agreement is best for mid-range velocities $v=0.2,0.3,0.4$, but we do
not expect our results to be quantitatively accurate for $v=0.1$.
\par

%ffffffffffffffffffffffffffffffffffffffffffffffffffffffffffffffffffffffff
\begin{figure}
\hspace{2.5cm}
%%%\epsfxsize=12cm
%%%\epsfbox{compare.deflect.minus.1.eps}
\caption{A comparison of scattering angles calculated by Myers et al.
to values calculated using the moduli space approximation when
$\l=-\frac{1}{16}$. The impact velocities shown are $v=0.1$ (+)
(note there are few points here because of vortex capture) and $v=0.2$ (o).}
\label{fig:compare.deflect.minus.1}
\end{figure}
%fffffffffffffffffffffffffffffffffffffffffffffffffffffffffffffffffffffffff

%ffffffffffffffffffffffffffffffffffffffffffffffffffffffffffffffffffffffff
\begin{figure}
\hspace{2.5cm}
%%%\epsfxsize=12cm
%%%\epsfbox{compare.deflect.minus.2.eps}
\caption{As above, with $v=0.3$ (x), $v=0.4$ (*).}
\label{fig:compare.deflect.minus.2}
\end{figure}
%fffffffffffffffffffffffffffffffffffffffffffffffffffffffffffffffffffffffff

A summary of the scattering angles versus impact parameters is shown in
Fig. \ref{fig:deflect.summary}, where the critical coupling
result is also marked.

%ffffffffffffffffffffffffffffffffffffffffffffffffffffffffffffffffffffffff
\begin{figure}
\hspace{2.5cm}
%%%\epsfxsize=12cm
%%%\epsfbox{deflect.summary.eps}
\caption{Summary of the scattering angles computed for
$\l=0$ (dashed line), $\l=\frac{1}{8}$ (above dashed line)
and $\l=-\frac{1}{16}$ (below dashed line).}
\label{fig:deflect.summary}
\end{figure}
%fffffffffffffffffffffffffffffffffffffffffffffffffffffffffffffffffffffffff

\section{Stability of bound orbits}
In this section we would like to consider the long term behaviour of a
bound system of two vortices, from a classical and
quantum mechanical point of view in the moduli space approximation.
The radial equation of motion for the two vortex system, expressed in the
coordinates $(\s,\theta)$ used in the previous section, can be
rewritten using the conserved energy $E=T+V$ (for convenience we
normalise $V$ here by subtracting $V(\infty)$ from it) and angular momentum
$h=\s^{2}F^{2}\dot{\theta}$ as
\eq
2\pi F^{2} \dot{\s}^{2} = E - \l V(\s) - \frac{2\pi
h^{2}}{F^{2}\s^{2}} \equiv E - w(\s,h)
\eqe
Thus the problem is a slightly  modified version of a particle in a
potential well. For $h \neq 0$, the metric radial
function $F$ works effectively to reduce the angular momentum
at smaller $\s$ and therefore increases the strength of the
centrifugal barrier as $\s \rightarrow 0$.
\par
When $\l<0$ and $h=0$, scattering with $E>0$ produces the familiar
$90^{\circ}$ behaviour. Motion with $\l V(0) < E <0$
produces a ``breather'' state, with the vortices separating to a
maximum distance, then returning and scattering through $90^{\circ}$
in each cycle. For $h \neq 0$, analysis of the function $w(\s)$
shows that for $h^{2} < 1.14|\l|$ , $w(\s,h)$ has two turning points,
the minimum at radius $\s<2.70$, the maximum at radius $\s>2.70$. This
implies the existence of trapped orbits.
For $h^{2} > 1.14|\l|$, no orbits are possible. Fig.
\ref{fig:effective.potential} illustrates the effective
potential $w(\s,h)$ for a range of values of $h$.

%ffffffffffffffffffffffffffffffffffffffffffffffffffffffffffffffffffffffff
\begin{figure}
\hspace{2.5cm}
%%%\epsfxsize=12cm
%%%\epsfbox{effective.potential.eps}
\caption{The effective potential $w(\s,h)$ for $h \in [0,0.1]$ at
intervals of $0.01$.}
\label{fig:effective.potential}
\end{figure}
%fffffffffffffffffffffffffffffffffffffffffffffffffffffffffffffffffffffffff

\par
The minimum of $w(\s)$ corresponds to a stable circular orbit in our
approximation. In the classical field theory, energy
will be radiated from the system, and the vortices will
spiral into a stable winding number two vortex. However,
this decay might not occur in the corresponding quantum
field theory. The available energy of the system is proportional to $\l$,
whereas the mass of the photon (smaller than the Higgs mass for our
range of parameters) is proportional to the gauge coupling $e$, scaled
to $e=1$ here. Hence if $\l$ is sufficiently small, there will not be
enough available energy for photons to be radiated.
\par
We do not attempt to analyse the quantum field theory here, but an
alternative method suggests itself: treating the zeros of the Higgs
field as quantum mechanical particles of mass $m$, and using
Schr\"odinger's equation to give an approximate quantisation.
For the wavefunction $\psi$, the time-independent Schr\"odinger
equation is
\eq
\frac{1}{\s}\frac{\del}{\del\s}(\s \frac{\del\psi}{\del\s}) +
\frac{1}{\s^{2}} \frac{\del^{2}\psi}{\del\theta^{2}} +
\frac{2m}{\hbar^{2}} F^{2}(E- \l V)\psi = 0
\eqe
We separate of variables, by setting
\eq
\psi = e^{\frac{il\theta}{\hbar}}R(\s)
\eqe
where $l=2n\hbar \, , \, n \in \integers$ as
$\theta \in [0,\pi)$. We can then reduce the
resulting equation for $R(\s)$ to a standard form by the
substitution $R=\chi\s^{-\half}$ giving
\eq
\chi^{\prime \prime} + \frac{2m}{\hbar^{2}} F^{2}(E-U)\chi = 0
\eqe
where
\eq
U(\s,n) = \frac{m\l}{\pi}V +
\frac{\hbar^{2}}{2mF^{2}\s^{2}}(4n^{2} - \frac{1}{4})
\eqe
The wavefunction decays exponentially in regions where $E<U(\s,n)$
(if the vortex mass is large, as is typically the case for an artifact
of symmetry breaking, the spectrum of allowed bound energy states
is dense, and the system behaves semiclassically. The
wavefunction is confined to the close neighbourhood of regions
where $E>U(\s,n)$. However, we need not assume that the vortex mass is
large here). As we are in two spatial dimensions,
there is always at least one energy level with $E<0$ (contrary to the
result for three dimensions) \cite{landau}.
For bound states with $E>0$ there is a small probability of
tunnelling through the potential barrier, allowing the
vortices to escape to infinity. States
with $E<0$ are confined to a neighbourhood of the classically accessible
region. The maximum energy available for decay of a bound
configuration with $E<0$ is $|\l| V_{depth}$, where
$V_{depth} = V(0)-V(\infty)$.
$m_{Higgs} = 1+ O(\l)$ is slightly smaller than $m_{photon} = 1$, but
this only changes the stability of an orbit to second order in $\l$.
Hence all configurations with $E<0$ are quantum mechanically stable if
\eq
|\l| < \frac{1}{V_{depth}} \simeq 0.31
\eqe

\vskip 1cm
\hskip -18pt
{\bf{Acknowledgements}}
\par
I am grateful to Robert Leese and Nick Manton for useful discussions.
In particular, I am grateful to Robert for the use of computer
facilities provided under grant GR/H67652 from the
SERC Computational Science Initiative.

\eject

\end{document}